\documentclass[aip,jcp,preprint,amsmath,amssymb,endfloats*]{revtex4}

\usepackage{graphicx}
\usepackage{dcolumn}
\usepackage{bm}
\usepackage{epstopdf}

\makeatletter 
\def\@dotsep{4.5} 
\makeatother

\newcommand{\be}{\begin{equation}}
\newcommand{\ee}{\end{equation}}
\newcommand{\bea}{\begin{eqnarray}}
\newcommand{\eea}{\end{eqnarray}}
\newcommand{\bdm}{\begin{displaymath}}
\newcommand{\edm}{\end{displaymath}}
\newcommand{\bit}{\begin{itemize}}
\newcommand{\eit}{\end{itemize}}
\newcommand{\ben}{\begin{enumerate}}
\newcommand{\een}{\end{enumerate}}

\begin{document}

\title{Lipid Flip-Flop  Driven Mechanical and Morphological Changes in Model Membranes} 

\author{Sanoop Ramachandran}
\email{sanoop@physics.iitm.ac.in}
\affiliation{Department of Physics, Indian Institute of Technology Madras, Chennai - 600 036, India}

\author{P. B. Sunil Kumar}
 \email{sunil@physics.iitm.ac.in}
\affiliation{Department of Physics, Indian Institute of Technology Madras, Chennai - 600 036, India\\
and MEMPHYS-Center for Biomembrane Physics, University of Southern Denmark, Odense, DK-5230, Denmark}

\author{Mohamed Laradji}
 \email{mlaradji@memphis.edu}
\affiliation{Department of Physics, The University of Memphis, Memphis, TN 38152-3390, USA\\
and MEMPHYS-Center for Biomembrane Physics, University of Southern Denmark, Odense, DK-5230, Denmark}

\date{\today}

\begin{abstract}
We study, using dissipative particle dynamics simulations,  the effect of 
active lipid flip-flop  on model fluid bilayer membranes.  We consider both cases of 
symmetric as well as asymmetric flip-flops.  Symmetric flip-flop leads to a steady state of 
the membrane with an  effective temperature higher than that of the equilibrium membrane and
an effective surface tension lower than that of the equilibrium membrane. 
Asymmetric flip-flop leads to transient conformational changes of the membrane 
in the form of bud  or blister formation, depending on the 
flip rate. 
\end{abstract}


\maketitle


\section{Introduction}

 Biomembranes are non-equilibrium structures 
due to the non-thermal energy contributions resulting from the activity 
of a wide variety of vicinal proteins. While the  phase behavior and morphology of lipid bilayer membranes 
have been the subject of  extensive amount of studies~\cite{mouritsen05},  most of these studies have been  on the  equilibrium 
properties of the membranes. This has changed during the last decade or so with investigations of the  effects, that active protein pumps have, on the 
undulations of lipid membranes, their morphology  and the normalization of their mechanical
constants ~\cite{manneville-99,ramaswamy-00,gautam-02,gov-04,gov-05,lomholt-06,giahi-07,schlomovitz-07}. 
Of particular importance is another  class of membrane-bound proteins, that actively
translocates phospholipids from one leaflet of a biomembrane to the other~\cite{daleke-03,devaux-06,pomorski-06}. 
The effects of these phospholipid translocators, on the mechanics and morphology of lipid membranes, 
has not received much attention. In  this article, we present a study 
of self-assembled lipid bilayers in the presence of active lipid translocation,
using dissipative particle dyanamics simulations.

Lipid synthesis in eukaryotic cells takes place almost exclusively on the cytosolic leaflet of the
endoplasmic reticulum (ER), which leads to an asymmetry in the lipid composition across the bilayer. 
In order to maintain a symmetric lipid density across the ER bilayer, nearly half of the 
newly synthesized lipids are rapidly translocated to the other leaflet~\cite{pomorski-06}. 
In contrast, the plasma membrane is marked by an acute asymmetry in the lipid composition. 
Indeed while phosphatidylcholine  and sphingomyelin are predominantly present in the exoplasmic leaflet,
phosphatidylserine and phosphatidylethanolamine  are mainly found in the cytosolic 
leaflet~\cite{alberts-cell}.  Maintenance of the symmetric lipid distribution in ER or the asymmetric lipid distribution in the plasma
membrane {\em cannot} be mediated by thermally induced lipid movements (also termed passive flip-flops) 
alone. Indeed, passive flip-flops of phospholipids are energetically unfavorable due to the 
large energy barrier, 20-50 kcal/mol,
associated with the translocation of the polar head group through the low dielectric permitivity hydrocarbon  core of the bilayer. Consequently,
the rate of passive flip-flops of phospholipids is extremely small, 
of the order of $10^{-5} {\rm s}^{-1}$~\cite{abreu-04,liu-05}, 
i.e. on average, a single lipid experiences a thermally induced flip-flop every 24 hours.
The lipid distribution across the bilayer is therefore actively maintained by a class of membrane-bound proteins known 
as phospholipid translocators~\cite{alberts-cell}. These include the adenosine triphosphate-dependent flippases
and floppases, and the energy-independent scramblases~\cite{daleke-03,pomorski-06,devaux-06}. 

Given the difficulties in purifying membrane-bound proteins,
most of lipid transcolators have not been identified. Moreover, 
the mechanism(s) of active flip-flop, mediated by
phospholipid translocators, remain elusive. Few models have been proposed as mechanisms for active 
phospholipid translocations~\cite{langley-79,kol-04}. In particular, 
Pomorski and Menon~\cite{pomorski-06} recently proposed a mechanism similar to that of swiping a magnetic card 
through a card reader. In this model, the hydrophobic head group of the flipped/flopped lipid 
(magnetic strip of the card) is shielded from the
hydrophobic environment of the bilayer, thereby facilitating its translocation across the bilayer. 
Recently, Sens~\cite{sens-04} investigated theoretically
the conformational response of an infinitely large membrane to a localized disturbance in the form of 
a localized transbilayer asymmetry in the lipid density. 
There, he found that this asymmetry may transiently  
lead to the formation of a bud-like invagination followed by its relaxation. 
In this paper, we present a
model for lipid translocation that is reminiscent of the magnetric swipe card
model ~\cite{pomorski-06}. The model is then investigated via large scale dissipative particle
dynamics simulations~\cite{sunil-mohamed-04,sunil-mohamed-05,laradji-06,yamamoto-02,shillcock-02}.  To our knowledge, the presented work is the first simulation study of the effect of active flip-flop on the mechanical and morphological
properties of lipid membranes.

\section{Bilayer Model}

In the  dissipative particle dynamics (DPD) model, for a self-assembled lipid bilayer in an explicit solvent used here, a lipid molecule is
modeled as a flexible amphiphilic chain of beads consisting of one ``head" ($H$) bead attached
to  three ``tail" ($T$) beads via Hookean spring bonds.  The solvent is modeled as 
single beads ($W$).  All particles have the  same mass $m$.  In this model, interactions between 
any two non-bonded particles, within  a range $r_0$, are soft and repulsive.  

The forces acting on particles are grouped into three categories: 
(i) conservative forces, (ii) dissipative forces, and (iii) random forces. The conservative 
force between any two particles is
\begin{equation}
{\bf F}_{ij}^{\left(C_1\right)}= \alpha_{ij}\omega(r_{ij})\hat{{\bf r}}_{ij},
\label{fconservative}
\end{equation}	
where $\alpha_{ij}$ is the interaction strength between particles 
$i$ and $j$, at respective positions ${\bf r}_i$ and ${\bf r}_j$,
${\bf r}_{ij}={\bf r}_i - {\bf r}_j$, and $\hat{{\bf r}}_{ij}=r_{ij}/|{\bf r}_{ij}|$.
Bonded particles belonging to a lipid also experience a conservative Hookean force given by
\begin{equation}
{\bf F}_{i,i+1}^{\left(C_2\right)}= -k\left(1-r_{i,i+1}/b\right)\hat{{\bf r}}_{i,i+1},
\label{fhookian}
\end{equation}	
where $k$ is the spring constant and $b$ is the preferred bond length.
The dissipative force between particles $i$ and $j$ is given by
\begin{equation} 
{\bf F}_{ij}^{\left(D\right)}=  -\Gamma_{ij}\omega^2(r_{ij})(\hat{{\bf r}}_{ij}\cdot{\bf v}_{ij})\hat{{\bf r}}_{ij},
\label{fdissip}
\end{equation}
where $\Gamma_{ij}$ is the dissipative strength for the pair $(i,j)$,
and ${\bf v}_{ij}={\bf v}_{i}-{\bf v}_{j}$ is their relative velocity. 
The random force between $i$ and $j$ is given by
\be
{\bf F}_{ij}^{\left(R\right)}= \sigma_{ij}(\Delta t)^{-1/2}\omega(r_{ij})\zeta_{ij}\hat{{\bf r}}_{ij},
\label{frand}
\ee	
where $\sigma_{ij}$ is the amplitude of the random noise for the pair $(i,j)$,
and $\zeta_{ij}$ is a random variable with zero mean and unit variance which
is uncorrelated for different pairs of particles and different time steps. Together,  the
dissipative and random forces act as a thermostat provided the fluctuation-dissipation 
theorem is satisfied. This yields to the following relation between $\Gamma_{ij}$ and $\sigma_{ij}$,
\be
\sigma_{ij}^2=2\Gamma_{ij} k_{\rm B}T,
\ee
where $k_{\rm B}$ is Boltzmann's constant and $T$ is the thermostat temperature.
In Eqs.~(\ref{fconservative}), (\ref{fdissip}) and (\ref{frand}), 
the weight factor, $\omega(r)$, is chosen as
\be
\omega(r)=
\begin{cases} 1-r/r_0 & \text{for $r$ $\leq$ $r_0$}\\
0& \text{for $r$ $>$ $r_0$,}
\end{cases}
\ee
where $r_0$ is the interactions cutoff.
The particles trajectories are obtained by solving Hamilton's equations using the 
velocity-Verlet integrator~\cite{ilpo-00}. In the simulation, $r_0$ and $m$ set the scales for length and mass, 
respectively. $k_{\rm B}T$ sets the energy scale. The time scale is given by  
$\tau=\left(mr_0^2/k_{\rm B}T\right)^{1/2}$. 
The numerical value of the amplitude of the random force 
is considered to be the same for all pairs and is given by 
$\sigma_{ij}=\sigma=3.0\left(k_{\rm B}^3T^3m/r_0^2\right)^{1/4}$,
and the fluid density $\rho=3.0r_0^{-3}$. The amplitudes of the conservative 
force are chosen to be $\alpha_{HH}=\alpha_{TT}=\alpha_{WW}=\alpha_{WH}=25k_{\rm B}T/r_0$ and 
$\alpha_{WT}=\alpha_{HT}=200k_{\rm B}T/r_0$. In Eq.~(\ref{fhookian}), the spring constant $k=100k_{\rm B}T$ and 
$b=0.45 r_0$.
The time step is chosen to be $\Delta t=0.01\tau$.   
The flat bilayer is initially constructed parallel to the $xy$ - plane and placed in the middle of the simulation box.  It is then allowed to equilibrate until its normal fluctuations attain saturation.  The total number of lipids used is 16,000 in a    
simulation box with size $L\times L \times L_z=(86 \times 86 \times 40)r_0^3$ for the case of symmetric flip-flops
and $(80\times 80 \times 46)r_0^3$ for the case of asymmetric flip-flops.  The system was subject
to periodic boundary conditions in all three directions. 

\section{Flip-Flop Scheme}

\begin{figure}[!ht]
\includegraphics[scale=0.5,angle=-90]{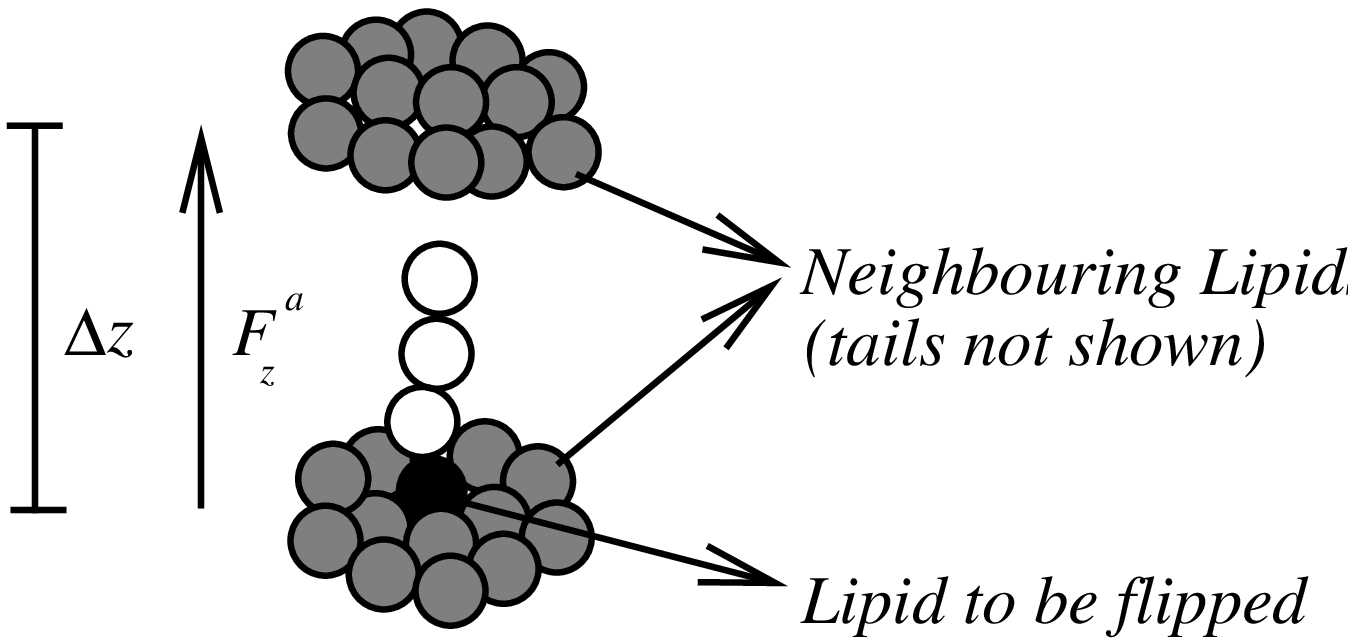}
\caption{Flippase complex corresponding to the fictitious cylinder containing the lipid to be flipped.  
The lipid to be flipped is shown explicitly. However, only head beads of the other lipids in the 
flippase complex are shown.} 
\label{fig:algo}
\end{figure}

We use the  following two-step scheme for ``flippase'' action: 
(i) formation of a complex and (ii) translocation of a lipid from one leaflet to another.  
A lipid to be flipped is randomly selected from one of the two leaflets
(refer to Figure ~\ref{fig:algo}, for a schematic representation of the lipid to be flipped and
the surrounding lipid molecules).  Around this selcted lipid a fictitious cylinder is drawn which
spans both leaflets of the bilayer.  A flippase complex is then defined as the set of all lipids
inside this fictitious cylinder. 
The next step involves the action of a time-dependent flipping force, ${\bf F}^{a}(t)=F_z^{a}(t)\hat{z}$, 
on the head bead of the selected lipid, so as to translocate 
it to the opposite leaflet.  This force, $F_z^{a}(t)$,  acts in the direction normal to the
plane of the bilayer, and its magnitude is given by  $F_z^{a}(t)=G\Delta z(t)$,
where $\Delta z(t)$ is the distance between the head bead of  the lipid being translocated and the 
average $z$-position of all the lipids, in the complex, in the opposite leaflet. This means that  during the translocation process, the amplitude of the
flippase force decreases continuously with time.
In order to conserve momentum within the fictitious cylinder,  
a force $-{\bf F}^{a}(t)/(N_c-1)$, where $N_c$ is the number of lipid molecules
in the fictitious complex, is concurrently applied onto the head beads of all other  lipids within the flippase complex. 
 During the translocation process of the selected lipid,
the head-tail repulsion of the selected lipid, with the other surrounding lipids in the
membrane, is ``screened" by temporarily choosing its amplitude $\alpha_{HT}=\alpha_{TT}$. This algorithm is therefore in-line
with the recent swipe card model~\cite{pomorski-06}.

\begin{figure}[!ht]
\includegraphics[scale=0.4]{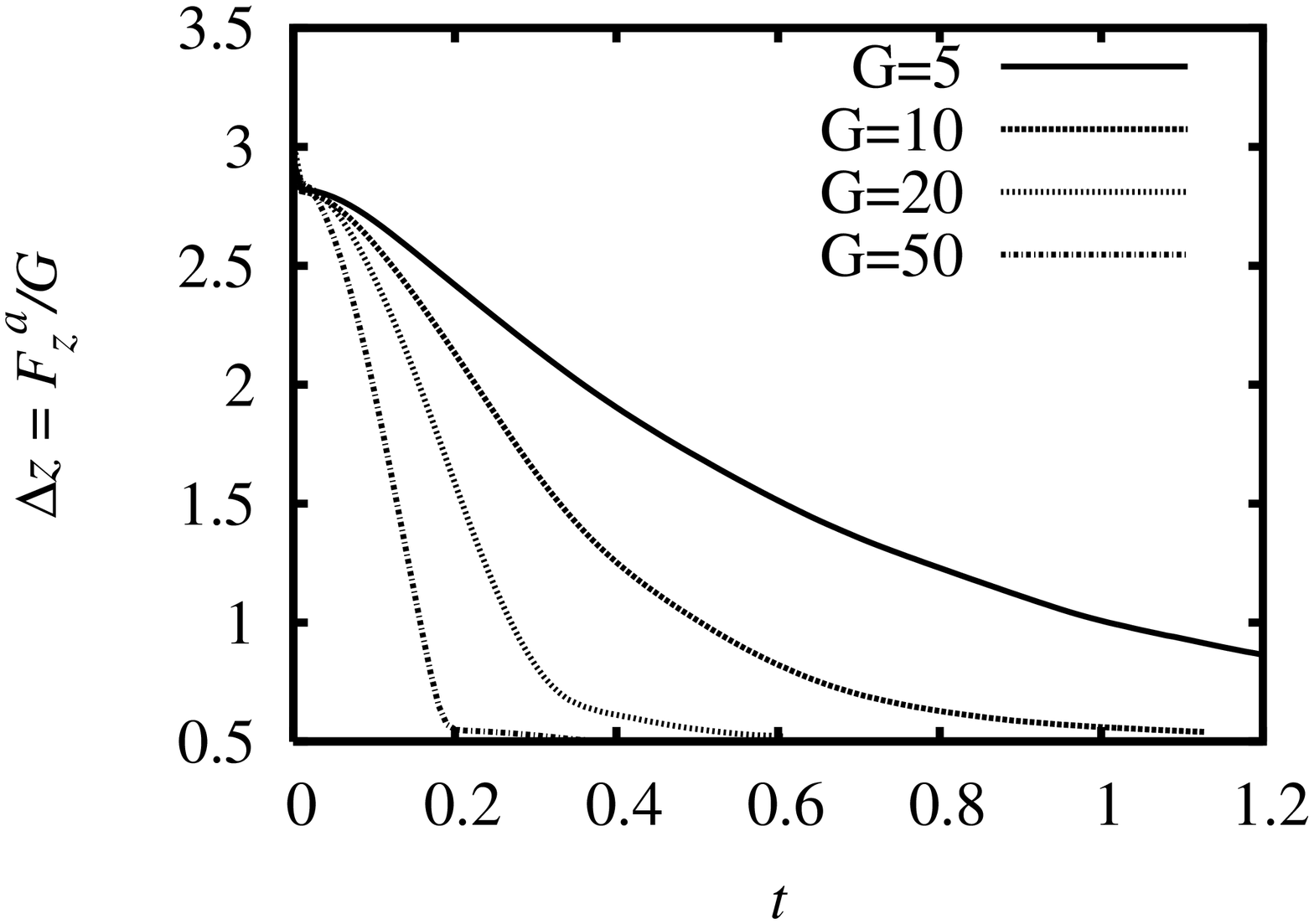}
\caption{Normalized amplitude of the translocation force, ${F_z^{a}}(t)/{G}$, vs. time for four
values of the force magnitude, $G$. The data has been averaged over all translocated lipids 
and over time during steady state.}
\label{fig:fvst}
\end{figure}

In Figure ~\ref{fig:fvst} the magnitude of the applied force, $F_z^a$, 
normalized by $G$, is shown as a function of time,
for the selected lipid as it translocates through the bilayer, mimicking the action of a flippase.  
This figure depicts that the translocation time scale decreases as the amplitude, $G$, of the driving force
is increased.  
In the remaining of the article, all results discussed are based on the case of $G=10k_{\rm B}T/r_0^2$, 
for which the typical time taken for flipping a lipid from one layer is about $\tau$.  During each time step,
a number of lipids are selected at random. Flips are attempted with a probability, $P_{\rm flip} $, 
if the selected lipid is not already part of an active flipping complex.  
The success {\it flip rate} is measured by  counting the number of lipids that reach the opposite leaflet 
in every time step. The flip and flop probabilities are respectively given by,
\be 
P_{\rm flip}  =  \frac{1}{1 + \exp\left[-A\left(N_u-N_d-C\right)\right]}, \label{eq:probability} \\
\ee
and 
\be 
P_{\rm flop}  =  \frac{1}{1 + \exp\left[-A\left(N_d-N_u+C\right)\right]}, \label{eq:probability} \\
\ee
where $N_u$ and $N_d$ are the numbers of lipid particles in the upper and lower leaflets, respectively.
In Eq.~(\ref{eq:probability}), $C$ controls the steady state number difference, $\Delta N=N_u-N_d$, 
and $A$ is constant which fixes the width of the
distribution of  $\Delta N$. Having developed an active translocation algorithm, 
we now focus on the effect that active flip-flop has on flat membranes.  We consider both cases of 
symmetric flip-flop, $C=0$, and asymmetric flip-flop, $C\ne 0$.

\section{ Symmetric Flip-Flop}

Active flip-flop is labeled symmetric, if on average, the number of up-down translocations (flips)
is equal to the number of  down-up translocations (flops).
At the beginning of the simulations,  we start with a flat bilayer with exactly the same number of lipids 
in the two leaflets. We note that the rate of thermally induced flip-flops in this model is practically zero, 
in accord with experiments.  
Using the Irving-Kirkwood formalism~\cite{irving-50}, we calculated the lateral and normal 
components of the pressure tensor along the $z$-axis,
and averaged over the $xy$-plane~\cite{schofield-82,goetz-98}.
The tension of the membrane and its bending modulus are extracted from the structure factor of the out-of-plane
fluctuations of the membrane height~\cite{sunil-mohamed-05}.

\subsection{Membrane Tension and Bending Rigidity}
 
\begin{figure}
\includegraphics[scale=0.4]{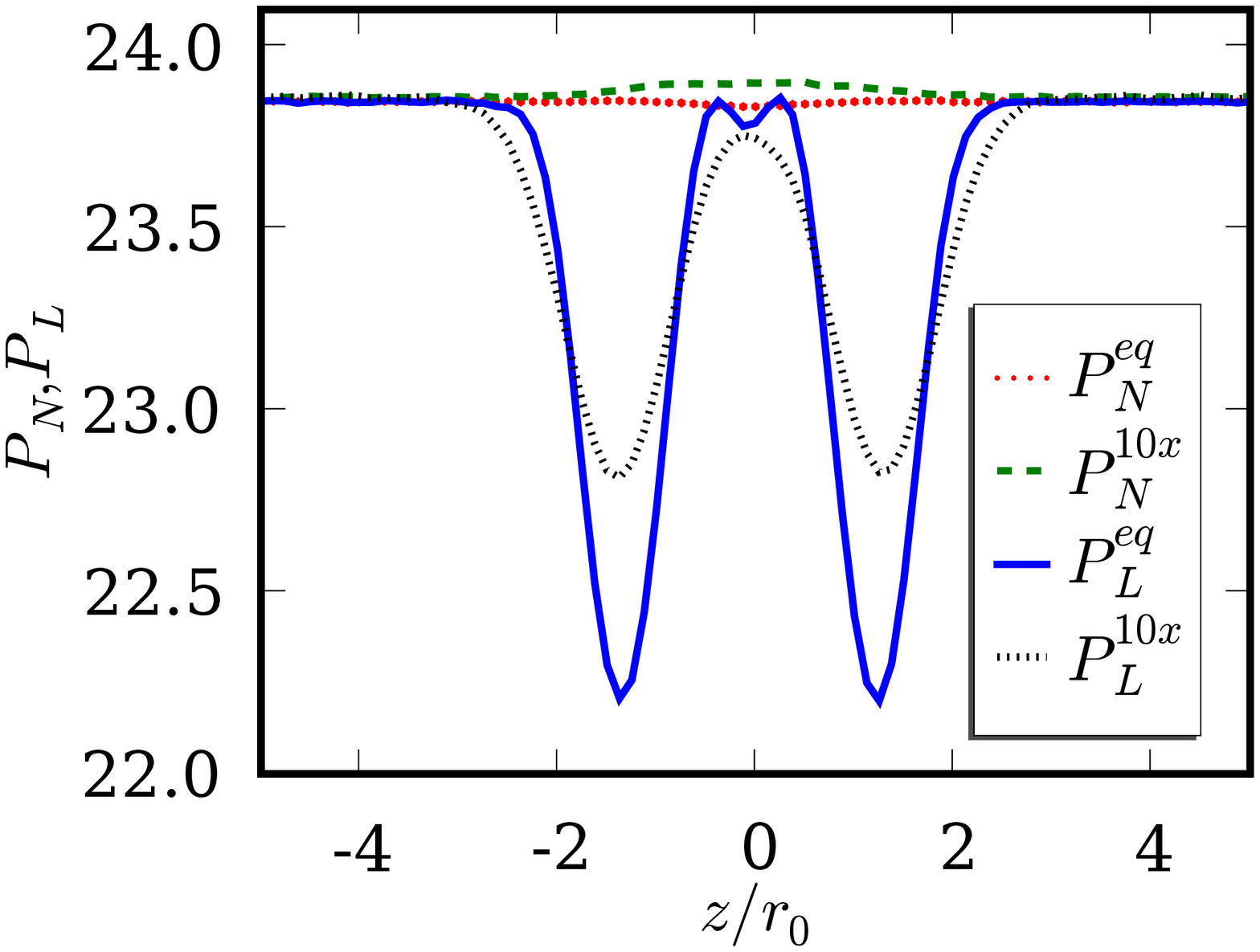}
\caption{Normal pressure $P_N$  and lateral pressure $P_L$ as a function of $z$ (both equilibrium and with an attempted flip rate of 10 flips per $\Delta t$) for a system with $L=86r_0$.}
\label{fig:pnpl}
\end{figure}

The average orientation of the layer normal is taken to be along the $z$-axis.  
The steady-state profiles of the normal  pressure $P_N(z)$ and the lateral pressure $P_L(z)$ along the bilayer normal
are calculated from the pressure tensor using the Irving-Kirkwood formalism~\cite{irving-50}.   
In Figure ~\ref{fig:pnpl}, the normal and lateral pressure vs. $z$  of equilibrium membranes are compared with that of membranes  active symmetric flip-flop.  The  rate of flipping is  $10$ flips/$\Delta t$ over a membrane with projected 
area $86r_0\times 86r_0$.
Although the details of the  lateral pressure profile is dependent on the model used for 
the lipids, one observes that the flip-flop activity 
increases the lateral pressure in the two leaflets while the normal pressure is only weakly affected. 
As a result, active flip-flop reduces the effective tension on the membrane.  

\begin{figure}[!ht]
\includegraphics[scale=0.4]{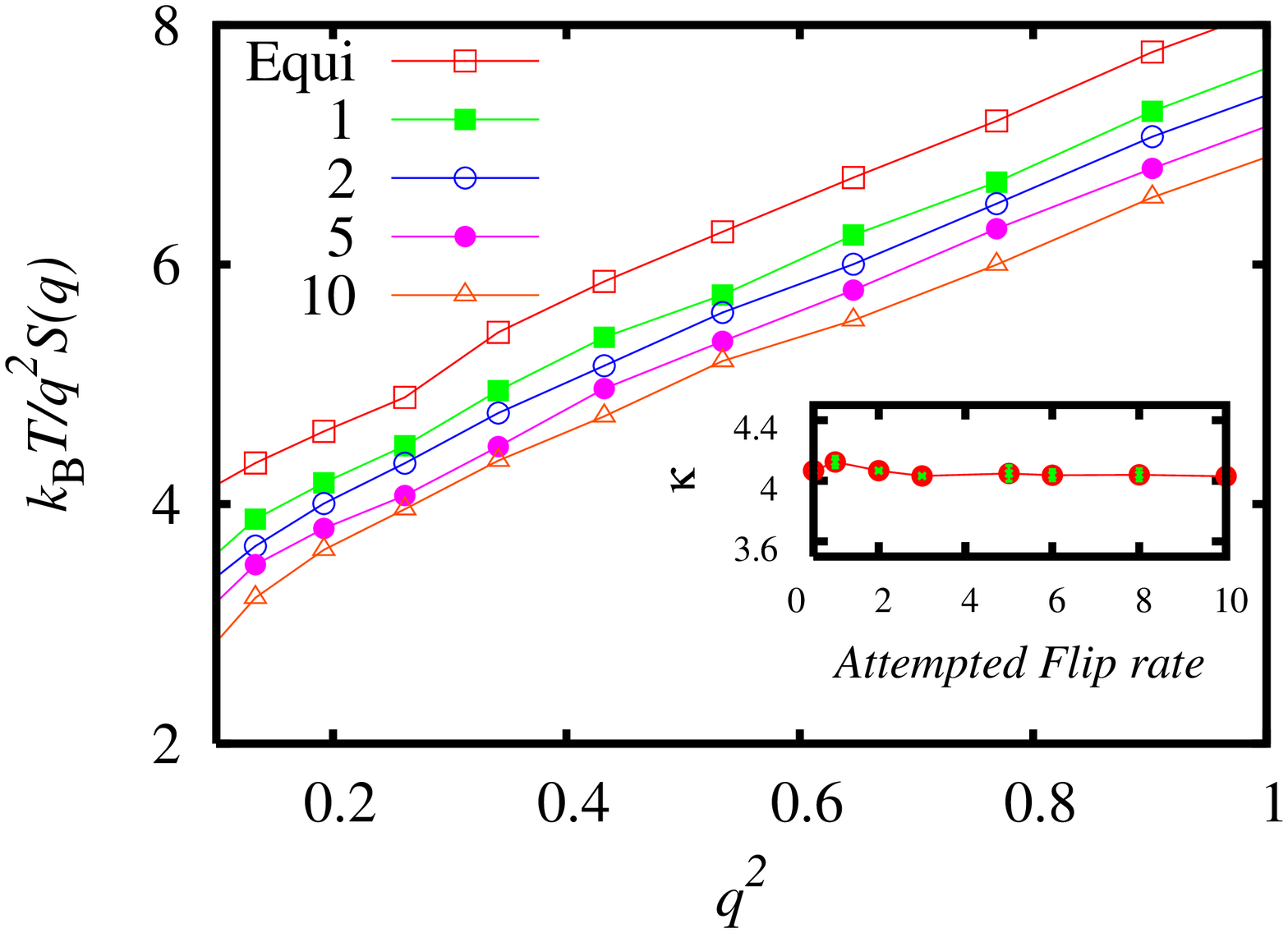}
 \caption{Structure factor vs. $q^2$ for different values of attempted flip rates. 
 Different symbols correspond to the cases of
 equilibrium ($\square$), 
1 flip$/\Delta t$ ($\blacksquare$), 2 flips$/\Delta t$ ($\circ$), 5 flips$/\Delta t$ ($\bullet$),
and 10 flips$/\Delta t$ ($\triangle$).
The simulations are performed on a system containing 16000 lipids with $L=86 r_0$ and $k_{\rm B}T=1.0$.}
\label{sigma-kappa}
\end{figure}

By defining a height field $h(x,y)$,  which represents the position  along the $z$-axis 
of the  bilayer mid-plane at a point $(x,y)$,  and its Fourier transform $\tilde{h}({\bf q})$ where ${\bf q}=(q_x,q_y)$, we calculate   
the circularly averaged structure factor 
$S(q) =  \langle | \tilde{h}({\bf q}) |^2 \rangle/{L^2}$, where
$q=\left(q_x^2+q_y^2\right)^{1/2}$.  
The long wavelength deformations of a lipid membrane from its mean planar conformation are well described by the
Helfrich Hamiltonian ~\cite{helfrich-73},   ${\cal H} [h(x,y)] = \int  dx dy \left[ \frac{\gamma}{2} (\nabla h)^2 + \frac{\kappa}{2} (\nabla^2 h)^2 \right]$,
where $\gamma$ is the membrane surface tension and $\kappa$ is its bending modulus. 
The equipartition theorem of this model yields a structure factor,
\be
S(q)=\frac{k_{\rm B} T}{\gamma q^2 + \kappa q^4}.
\ee
Hence, by plotting $k_{\rm B}T/q^2 S(q) $ as a function of $q^2$, one extracts 
the tension on the membrane,
$\gamma$, from  the intercept with the vertical axis 
and the bending modulus, $\kappa$, from the slope at small wavevectors.   
This is shown in Figure ~\ref{sigma-kappa} for the cases of equilibrium and steady state flip-flop with varying flip-flop rates. Figure ~\ref{sigma-kappa} shows that as the rate of flip-flop is increased, 
the intercept of $k_{\rm B}T/q^2 S({\bf q})$
is shifted to lower values, implying a reduction in the tension of the membrane. This is in 
line with the results from the lateral and normal pressure shown in Figure ~\ref{fig:pnpl}.  
However, Figure ~\ref{sigma-kappa} shows that the slope of  $k_{\rm B}T/q^2 S(q)$ 
is independent of
the flip-flop rate, implying that for the flip rates considered in our simulation,
there does not seem to be any significant effect on the membrane bending modulus due to active flip-flop.

We thus conclude that symmetric active flip-flops leads to an increase in the fluctuations of the membrane,
which manifests itself in a decrease of the
surface tension of the membrane.

\section{ Asymmetric Flip-Flop}

We now consider the effect of asymmetric flip-flop. In particular, we focus on the
 case where only lipids from the bottom leaflet
are actively flipped to the upper leaflet.  This is implemented  
by assigning  a non-zero value for  the constant $C$ in the expression for the probability $P_{\rm flip}$ 
in Eq.~(\ref{eq:probability}).  
The value of $C$ is kept large, equal to $10^4$, such that the rate of accepted flips can be 
taken as a constant during simulation time.  Flipping is restricted to a small square 
area ($l \times l$), termed the active region, in the central region of the membrane. 
This mimics the effect of flips due to localized 
flippases in a small region of a biomembrane.   
The effect of asymmetric flipping is then investigated as a function of $l$  and 
the total number of flips per unit time ($\nu$) . $l$ is varied between $10r_0$ to $50r_0$ and 
the flip rate  is varied from $\nu=0.9 \,\tau^{-1}$ to $\nu=20 \, \tau^{-1}$. 

Asymmetric flip induces a finite difference in the lipid number densities, $\Delta s =s_u-s_d$, where
$s_u$ and $s_d$ are the lipid number densities per unit of area,
of the upper and lower leaflets, respectively. Furthermore,
$\Delta s$ increases with time as active flip proceeds. Therefore, asymmetric flip is characterized by 
an absence of steady state, and if maintained, leads to an instability of the membrane,
in contrast to the case of symmetric active flip-flop. 
We found that asymmetric active flipping lead to the formation of two major transient morphologies corresponding to
either buds or blisters, depending on size of the flip area and the flip rate.

\begin{figure}
\includegraphics[scale=0.6]{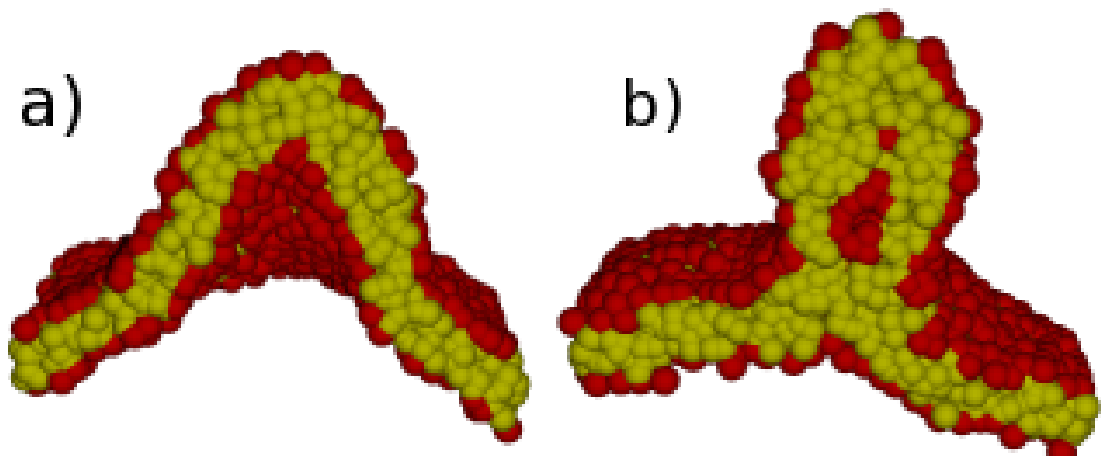}
\caption{Bud Formation (only a small section of the membrane, with a cut through the budding region,  is shown). a) Initial bending of both the leaflets and b) neck formation which eventually leads to a bud. The flip rate and size of the active region are  $\nu=0.9\tau^{-1}$ and $l=10$ respectively.}
 \label{bud_formation}
\end{figure}

When the flip rate is low, $\nu <2\,\tau^{-1}$, we found that buds form. This is shown   in Figure ~\ref{bud_formation}.  In this case, a full bud is formed after about $800\tau$. 
Once formed, the bud remains stable even after the flipping is stopped, and will relax back only in
the time scale set by passive flips, which is much larger than the simulation time.  The stability of
the buds here has to do with the finite size of the membrane~\cite{svetina-89,miao-94}.  For an infinite membrane,  the buds have a finite lifetime set by the relaxation of the in-plane density~\cite{sens-04}.

\begin{figure}
\includegraphics[scale=0.6]{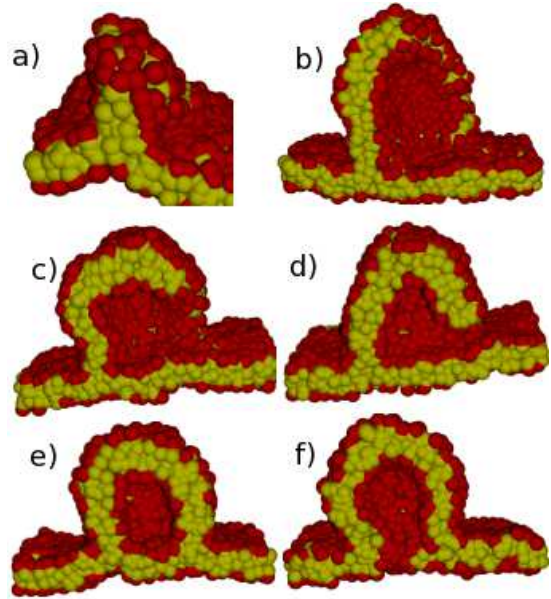}
\caption{Blister Formation (only a small section of the membrane, with a cut through the blister,  is shown).  a) Initial protrusion b) sheet formation c-e) fusion of sheet with membrane and f) pore formation. Note that the edge of the blister, clearly visible in in (b), moves towards the cut plane  in (b) to (e). The flip rate and size of the active region are  $\nu=2.5\tau^{-1}$ and $l=30$ respectively. }
\label{blister_formation}
\end{figure}

\begin{figure}
\includegraphics[scale=0.6]{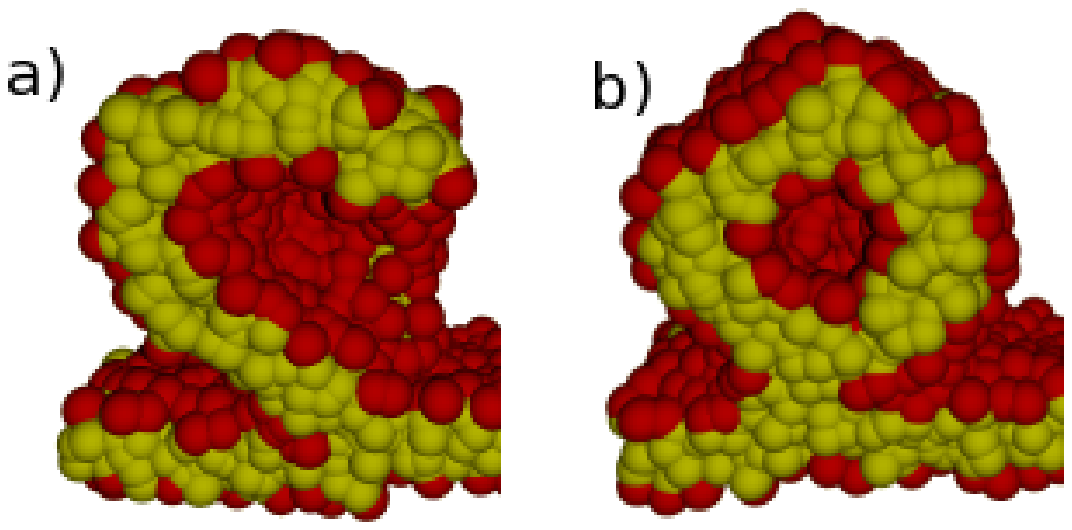}
\caption{In its early stage blisters can also relax by folding into itself, forming a bulb-like structure.} 
\label{blister_stalk}
\end{figure}

When lipids are flipped at a relatively high rate, we observed the formation of blister structures 
in the active region. In the case with flipping  rate $\nu=2.5 \tau^{-1}$, it takes about $300\tau$
for a blister to form. 
In Figure ~\ref{blister_formation},  we depict snapshots showing the main stages of blister formation at this flip rate.
The relatively high active flip rate and slow diffusion of the lipids 
results in high excess of  lipids in the upper leaflet. This leads to the detachment of the  upper leaflet from the lower leaflet, resulting in  a protrusion 
that is reminiscent of a cylindrical micelle connected to the membrane. This is depicted  in Figure ~\ref{blister_formation}(a). 
The cylindrical micelle then grows into a sheet-like structure still connected to the membrane, as
shown in Figure ~\ref{blister_formation}(b). If the process of active flipping is further continued, we found that the
membrane becomes unstable. To avoid a destabilization of the membrane, we consider only the cases where 
active flipping  is stopped once the sheet-like structure (Figure ~\ref{blister_formation}(b)) is formed. The  relaxation of the blister depends on 
the stage at which active flipping is halted. If active flipping is stopped during the early stage of blister formation , shown in  Figure ~\ref{blister_formation}(a),  the blister relaxes back into the membrane. At late stages, when the sheet is well formed (see Figure ~\ref{blister_formation}(b)) , in order to reduce the  high edge energy,  the blister curves forming a hook-like structure  shown in Figure ~\ref{blister_formation}(c). Eventually, the sheet closes on to the membrane,  forming a hemifusion state, 
reminiscent of that observed during the intermediate stages of vesicles fusion~\cite{shillcock-05}.  This is elucidated in the series of snapshots shown in Figure ~\ref{blister_formation}(b-e).
Eventually, the diaphragm  in the hemifusion state ruptures into a pore as shown in Figure ~\ref{blister_formation}(f). If active flip is stopped at an intermediate stage between  those shown in 
Figure~\ref{blister_formation}(a) and (b), the blister relaxation is arrested at the hemifusion state  shown in Figure ~\ref{blister_stalk}(b).

\section{Conclusion}
The thickness of the bilayer obtained in the simulations is $4 r_0$. Comparing this with the 
thickness of lipid bilayers, $5\ {\rm nm}$, we obtain  $r_0\approx 1.25 ~{\rm nm}$.  
In phospholipid bilayers in the fluid phase, the diffusion coefficient of a lipid is typically
$D\sim 10^{-12} {\rm m^2/s}$~\cite{alberts-cell}.  Comparing this with the diffusion 
coefficient for lipids obtained in the current simulations, one estimates the DPD time unit, 
$\tau\approx 0.2 {\rm \mu s}$.
For the case of $G=5 k_BT/r_0^2$, we then find that the average duration 
of a single flip is about $2 {\rm \mu s}$.  Considering that the time scale for  typical 
protein conformation change ranges from milliseconds  to seconds,  
to make contact with real systems, one has  to use smaller values of $G$. 
However, for practical reasons we use $G=10 k_BT/r_0^2$. The typical simulation run in the present work is  
about $200 {\rm \mu s}$ corresponding to a diffusion length of approximately $15 {\rm nm}$.  
Although the lipid density equilibrates very fast, the budding event is a result of the global 
asymmetry in the lipid number.

 In the case of an infinite membrane the formed buds will disappear on time scales larger than
diffusion time. However in finite systems  the resulting  area difference of the two leaflets 
leads to buds that are stable within the time scale  set by  passive flip-flop rates~\cite{svetina-89,miao-94}.

The initial tension on the membrane also plays a role in determining the critical flip rate for  bud and blister formation. In the case of membranes with fixed projected area, asymmetric flipping  
from the bottom leaflet to the upper leaflet  leads to an asymmetry in the area per lipid in the two leaflets of 
the membrane. This in turn leads to an increase in  the lateral tension of the bottom leaflet while that of the upper leaflet decreases. In order to equalize the area per lipid in the two leaflets, 
the membrane buckles, which will further increase the tension in both the leaves.  
Beyond a critical tension, the bottom leaflet will either  rupture or  decouple from the upper leaflet resulting in a blister. 
This also means that the flip rate at which blisters start to form should decrease with increasing  tension of the initial 
equilibrated membrane. We confirm this in our simulations.
 
In conclusion, we have presented a DPD model to study active flipping and its effects  on a
fluid  bilayer membrane. We find that symmetric flip-flop  results in a reduction 
in the tension of the membrane without much effect on its bending modulus.  Asymmetric flip-flop
results in non-equilibrium structures  depending on the  lipid flip rate.  
Slow flip rates involves membrane curvature and bud formation, whereas fast
flip rates induces blister formation.

\section*{Acknowledgements}
PBSK acknowledges DST India for financial support.  
ML acknowledges the financial support through a grant from the Research Corporation (Award No. CC6689). 
MEMPHYS is supported by the Danish National Research Foundation.


\listoffigures
\end{document}